# Shear of the vector potential in the Aharonov-Bohm effect

Keith J. Kasunic

*University of California at Irvine, Irvine, CA 92697*

**Abstract**

The Aharonov-Bohm (AB) effect is now largely considered to be a manifestation of geometric phase. However, by decomposing the vector-potential gradient tensor into divergence, curl, and shear components, we isolate a field/charged-particle interaction that is not dependent on local electric and magnetic fields. We show that a local shear field provides a velocity-dependent, dynamic-phase interaction in the AB effect whose predictions are consistent with all known classes of AB experiments, including interference fringe shifts, the absence of time delays along the direction of propagation, and the possibility of lateral forces.

**PhySH:** Aharonov-Bohm effect, quantum interference effects, electromagnetic interactions, geometric and topological phases

## I. Background

The Aharonov-Bohm (AB) effect – where electron-interference fringe positions are shifted based on the presence of a non-local magnetic field [1]-[10] – is now largely considered to be a manifestation of geometric phase [8]. What is still not clear, however, is whether this geometric-phase solution is one of necessity or one of utility. As illustrated by the Foucault pendulum, for example, geometric- and dynamic-phase descriptions can both lead to the same predictions of experimental results.

While there have been a number of force-based interpretations discussed in the literature for more than 60 years – e.g., Bohm's notion of the quantum potential [3], the collimated-beam scattering analyses of quantum particles by Shelankov [4] and Berry [5], the force-impulse derivations of Keating and Robbins [6], the non-local "force" measured by Becker *et al.* [7], and the induced **E**-fields of Boyer [11] – none have clearly identified what magnitudes, directions, and time scales might be involved in the AB phase shift $\Delta\phi_{AB} = e\Phi_B/\hbar$ that results in an electron-interference fringe shift in regions of space where there are neither electric nor magnetic fields. In this paper, we derive a velocity-dependent field/charged-particle interaction using a shear tensor whose components are determined by gradients in the vector potential **A**. This tensor represents a property of **A** whose components are not dependent on local electric or magnetic fields.

## II. Shear of the Vector Potential

In a recent paper, it was reported that the AB phase shift can be derived semi-classically from changes in the electron's wavelength between the upper and lower halves of an AB solenoid [9]. This shift can be viewed as a result of changes in the de Broglie wavelength, driven by a gauge-invariant transfer of electromagnetic field momentum $e\mathbf{A}$ to (and from) the electron's mechanical momentum $m\mathbf{v}$, where $m\Delta\mathbf{v} = -e\Delta\mathbf{A}$ from conservation of canonical momentum $\mathbf{p}_c = m\mathbf{v} + e\mathbf{A}$ in the azimuthal direction where $\partial A_\theta(r)/\partial\theta = 0$. The vector potential outside the solenoid has circulation (Fig. 1), but neither divergence ($\nabla\cdot\mathbf{A} = 0$ from use of the Coulomb gauge) nor curl ($\mathbf{B} = \nabla\times\mathbf{A} = 0$ from the absence of magnetic fields).

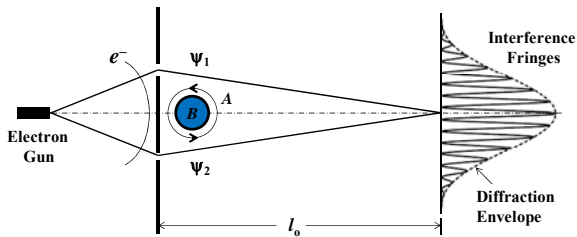

Figure 1 – Schematic of the Aharonov-Bohm (AB) two-slit interference experiment, where the magnetic field **B** is zero in the region outside the solenoid (shown in blue), yet the waves $\psi_1$ and $\psi_2$ for the electron $e^-$ recombine with different maxima and minima locations, depending on the magnitude of the magnetic vector potential **A**.

In Ref. [9], any forces responsible for the momentum transfer were not identified; to do so, we apply the results of Romano and Price [12] to introduce the gradient of the vector potential

$$\frac{\partial A_i}{\partial x^j} = \sigma_{ij} - \frac{1}{2}\varepsilon_{ijk}(\nabla\times\mathbf{A})_k + \frac{1}{3}\delta_{ij}(\nabla\cdot\mathbf{A}) \quad (1)$$

where $\varepsilon_{ijk}$ is the Levi-Ceviti symbol, $\delta_{ij}$ is the Kronecker delta, and $\sigma_{ij}$ are the components of the symmetric shear tensor [12]

$$\sigma_{ij} \equiv \frac{1}{2}\left(\frac{\partial A_i}{\partial x^j} + \frac{\partial A_j}{\partial x^i}\right) - \frac{1}{3}\delta_{ij}(\nabla\cdot\mathbf{A}) \quad (2)$$

which, with both the divergence and curl in Eqn. (1) equal to zero, uniquely determines the second-rank vector-potential gradient tensor $\nabla\mathbf{A}$.

For the AB effect, we determine **A** using the Coulomb gauge (giving $A_r = 0$) and Stokes' Theorem applied to the circulation of **A** around the solenoid

$$\oint\mathbf{A}\cdot d\mathbf{s} = \iint(\nabla\times\mathbf{A})\cdot d\mathbf{S} = \Phi_B \quad (3)$$

for a magnetic field **B** and magnetic flux $\Phi_B$ inside the solenoid. Symmetry then fixes the only component of **A** as the tangential $A_\theta(r) = \Phi_B/2\pi r$ for $r \geq R$ (solenoid radius) in cylindrical coordinates; Eq. (2) for the shear of the vector potential then gives

$$\sigma_{r\theta} = \sigma_{\theta r} = \frac{1}{2}\left(\frac{\partial A_\theta}{\partial r} - \frac{A_\theta}{r}\right) = -\frac{\Phi_B}{2\pi r^2} \quad (4)$$

which can be interpreted as an angular deformation of the field elements. In addition to the inverse-square variation with radius, we see that the shear depends only on the circulation of the vector potential outside the solenoid, i.e., the magnetic flux inside the solenoid.

Converting this shear of the vector-potential field to a force on an electron moving through it, a dimensional analysis shows that the shear of the field momentum $e\mathbf{A}$ has units of $mv$ per meter, or kg/sec; multiplying by the electron velocity gives us units of Newtons. We initially use the electron free-stream velocity $v_o$, giving a shear force $\mathbf{F}_\theta$

$$\mathbf{F}_\theta \sim ev_o\sigma_{r\theta}\hat{\mathbf{\theta}} = -ev_o\frac{\Phi_B}{2\pi r^2}\hat{\mathbf{\theta}} \qquad (5)$$

where the units of the shear $\Phi_B/2\pi r^2$ are that of magnetic field, giving a force that is similar in scalar form to the velocity-dependent "$evB$" Lorentz force. This is, of course, not a Lorentz force, as there is no curl of the vector potential in the region outside the solenoid; instead, we have found a velocity-dependent force on a charged particle due to the shear of the vector potential, whose radial distribution we can approximate using Eq. (5).

Going beyond this simple approximation, we next treat the electron not as a point mass with speed $v_o$, but as a velocity field $\mathbf{v}(r, \theta)$ containing a radial and polar-angle dependence on the geometric boundary conditions imposed by the solenoid. Just as with the hydrodynamic velocity field around a cylinder, we find this dependence using potential-flow solutions to Laplace's equation. Specifically, from the absence of divergence and curl of $\mathbf{A}$ outside the solenoid, where the electron mechanical momentum $\mathbf{p} = \mathbf{p}_o - e\mathbf{A}$ for a constant "free-stream" momentum $\mathbf{p}_o = p_o\hat{\mathbf{x}}$ away from the influence of the solenoid, we have

$$m\nabla \cdot \mathbf{v} = -e\nabla \cdot \mathbf{A} = 0 \qquad (6)$$

and

$$m\nabla \times \mathbf{v} = -e\nabla \times \mathbf{A} = 0 \qquad (7)$$

showing that the electron velocity field $\mathbf{v}(r,\theta) = \nabla\phi_v$ outside the solenoid is also solenoidal and irrotational. From the absence of curl, we can write $\nabla \times \mathbf{v} = \nabla \times \nabla\phi_v = 0$; from the absence of divergence, we have $\nabla \cdot \mathbf{v} = \nabla \cdot \nabla\phi_v = \nabla^2\phi_v = 0$. The velocity field can thus be found from a solution to Laplace's equation for a scalar velocity potential $\phi_v(r,\theta)$. For propagation through a source of clockwise velocity circulation $\Gamma_v$ around a cylinder of radius $R$, the velocity potential $\phi_v(r,\theta)$ is given by [13]

$$\phi_v(r,\theta) = v_o\cos\theta\left(r + \frac{R^2}{r}\right) + \frac{\Gamma_v}{2\pi}\theta \qquad (8)$$

from which we obtain

$$v_r(r,\theta) = \frac{\partial\phi_v}{\partial r} = v_o\cos\theta\left(1 - \frac{R^2}{r^2}\right) \qquad (9)$$

and

$$v_\theta(r,\theta) = \frac{1}{r}\frac{\partial\phi_v}{\partial\theta} = -v_o\sin\theta\left(1 + \frac{R^2}{r^2}\right) + \frac{\Gamma_v}{2\pi r} \qquad (10)$$

for $r \geq R$ and an angle $\theta$ measured in the conventional sense as counter-clockwise around the solenoid origin with respect to the positive $x$-axis (Fig. 2); from Eq. (7), the velocity circulation $\Gamma_v$ scales by a factor of $e/m$ from the vector potential circulation

$$\Gamma_v \equiv \oint \mathbf{v} \cdot d\mathbf{s} = \iint (\nabla \times \mathbf{v}) \cdot d\mathbf{S}$$
$$= -\frac{e}{m} \iint (\nabla \times \mathbf{A}) \cdot d\mathbf{S} = -\frac{e}{m} \Phi_B \quad (11)$$

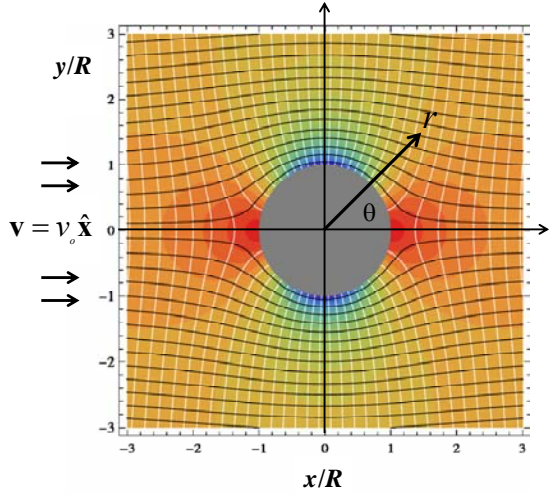

Figure 2 – Solutions to Laplace's equation for the electron trajectories with $\Gamma_v = 0$ show the influence of the solenoid geometry (in grey) on an electron incident from the left with velocity $\mathbf{v} = v_o \hat{\mathbf{x}}$. Adapted from

Note that $v_r = 0$ at $r = R$ for any azimuthal angle $\theta$, as required by the kinematic ("impenetrability") boundary condition for the radial velocity at the solenoid surface. While there is a clear analogy here with the potential flow of an ideal (inviscid, irrotational, incompressible) fluid around a cylinder, this is not a de Broglie-Bohm quantum hydrodynamics model, but a classical solution to Laplace's equation for an electron velocity field which has neither divergence nor curl.

To determine which of these velocity components applies to the tangential force in Eq. (5), we use the total derivative $d\mathbf{A}/dt$ for convective momentum transport, with canonical momentum conservation $d\mathbf{p}_c/dt = 0$ giving a force $md\mathbf{v}/dt = -ed\mathbf{A}/dt$, where

$$\frac{d\mathbf{A}}{dt} = \frac{\partial \mathbf{A}}{\partial t} + (\mathbf{v} \cdot \nabla)\mathbf{A} \quad (12)$$
$$= v_r \frac{\partial A_\theta}{\partial r} \hat{\boldsymbol{\theta}} - \frac{v_\theta A_\theta}{r} \hat{\mathbf{r}}$$

for $\mathbf{A} = A_\theta(r)\hat{\boldsymbol{\theta}}$, $\partial \hat{\boldsymbol{\theta}}/\partial r = 0$, and $\partial \hat{\boldsymbol{\theta}}/\partial \theta = -\hat{\mathbf{r}}$ when expanding the partial derivatives of the $\mathbf{v} \cdot \nabla$ operator (see Appendix A). We also assume a stationary field at all points ($\mathbf{E} = -\partial \mathbf{A}/\partial t = 0$) to remove the effects of any electric field $\mathbf{E}$. As the particle moves, it will see different values of the vector potential, changing its velocity via momentum transfer from the field momentum. Based on how quickly the particle moves at velocity $\mathbf{v}$ through the vector-potential gradient $\nabla \mathbf{A}$, the acceleration of the particle is thus calculated only from the convective derivative $(\mathbf{v} \cdot \nabla)\mathbf{A}$ determining the local change in field momentum.

While $d\mathbf{A}/dt$ has previously been used in its general form in the Euler-Lagrange derivation of the Lorentz force law [14], Eq. (12) is not an equation of motion, and is only used here to identify which velocity terms are associated with any possible force components in the AB effect. Comparing Eq. (12) with the tangential force in Eq. (5), we see that $v_r$ is the velocity component for the tangential shear. Using $v_r$ from Eq. (9) instead of $v_o$ in Eq. (5), and substituting for $\partial A_\theta/\partial r = -\Phi_B/2\pi r^2$, we have an expression for the shear force distribution around the solenoid

$$\mathbf{F}_\theta(r, \theta) = ev_r \frac{\partial A_\theta}{\partial r} \hat{\boldsymbol{\theta}} \quad (13)$$
$$= -ev_o \cos\theta \left(1 - \frac{R^2}{r^2}\right) \frac{\Phi_B}{2\pi r^2} \hat{\boldsymbol{\theta}}$$

From the total derivative in Eq. (12), we also have an expression for the radial (centripetal) force component required for the change in direction of electron momentum

$$\mathbf{F}_r(r,\theta) = -ev_\theta \frac{A_\theta}{r}\hat{\mathbf{r}} \quad (14)$$

$$= e\left[v_o \sin\theta\left(1 + \frac{R^2}{r^2}\right) - \frac{\Gamma_v}{2\pi r}\right]\frac{\Phi_B}{2\pi r^2}\hat{\mathbf{r}}$$

and the AB force components in Eqns. (13) and (14) can be written as an inner product

$$\mathbf{F}_{AB} = e\mathbf{v} \cdot \ddot{\boldsymbol{\sigma}}_{AB} \quad (15)$$

for an AB shear tensor $\sigma_{AB}$ whose entries $\sigma_{ij}$ from Eq. (2) are all zero except $\sigma_{r\theta} = \partial A_\theta/\partial r = \sigma_{\theta r} = -A_\theta/r$, as expected for a symmetric tensor. This result is gauge invariant in that any addition to $\mathbf{A}$ such as $\partial_\mu \chi$ leaves the *magnetic* field unchanged. However, while we can still choose whatever gauge is convenient for the geometry and time-dependence at hand, we no longer have the freedom to add any quantities to the vector potential within that gauge, as the shear tensor has imposed an additional constraint on what is otherwise an under-constrained problem. This is seen from the equality $\sigma_{r\theta} = \sigma_{\theta r}$, where additions to $\mathbf{A}$ are not allowed due to their effect on $\sigma_{\theta r} \sim A_\theta$.

A plot of the radial dependence of the tangential component $F_\theta$ at $\theta = 180$ degrees (solenoid leading edge) is shown in Fig. 3(a) with a maximum at $r/R = 2^{1/2}$, after which it asymptotically approaches $1/r^2$ at large $r$. The radial dependence of the radial force $F_r$ at $\theta = 90$ degrees is also shown; the velocity-circulation $\Gamma_v$ term [with $v_o = 0.6 \times 10^8$ m/s ($E_o = 10$ keV) and $\Phi_B \sim 10^{-15}$ Wb in Eq. (11)] is negligible in comparison with the $v_o\sin\theta$ component of the radial force $F_r$ in Eq. (14) over most of the field, and is set to zero in Fig. 3(b) where $\sin\theta = 1$ for the radial distribution.

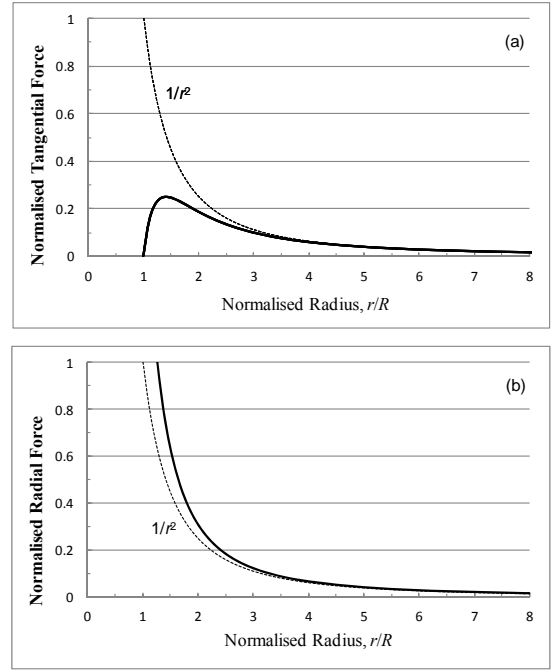

Figure 3 – (a) The normalized tangential force $F_\theta$ [the solid line = $(1 - R^2/r^2)/r^2$ from Eq. (13)] on an electron moving towards the solenoid at $\theta = 180$ degrees has a maximum at $r/R = 2^{1/2}$ and approaches $1/r^2$ at large $r$. (b) The normalized radial force $F_r$ at $\theta = 90$ degrees [the solid line = $(1 + R^2/r^2)/r^2$ from Eq. (14)] sets $\Gamma_v \approx 0$ for typical experimental conditions.

### III. Discussion

We can express the force distributions in Eqns. (13)-(14) in terms of their Cartesian components, allowing a general comparison with experimental results. Using standard coordinate transformations and again ignoring the circulation term $\Gamma_v$ in Eq. (14), we have

$$F_x(r,\theta) = F_r\cos\theta - F_\theta\sin\theta \quad (16)$$
$$\approx 2ev_o\sin\theta\cos\theta\frac{\Phi_B}{2\pi r^2}$$

and

$$F_y(r,\theta) = F_r\sin\theta + F_\theta\cos\theta$$
$$\approx ev_o\left[\sin^2\theta\left(1+\frac{R^2}{r^2}\right) - \cos^2\theta\left(1-\frac{R^2}{r^2}\right)\right]\frac{\Phi_B}{2\pi r^2} \quad (17)$$

giving a longitudinal force $F_x$ in the direction of propagation – with a $\cos\theta\cdot\sin\theta$ term which has a period of one-half that of the sine or cosine – and an angle-averaged force of zero over the upper or lower half of the solenoid ($\Delta\theta = \pi$) at any radius $r \geq R$. This is consistent with recent AB experiments by Becker and Batelaan showing no measureable time delay for the longitudinal propagation of an electron [15]. The physical interpretation is that the electron sees a positive net acceleration from the shear force over the top half of the solenoid, and a net deceleration over the bottom half, with no shift in its center-of-mass along the propagation axis.

For the lateral force $F_y$ given by Eq. (17), Becker et al. have recently published results showing an asymmetry in the AB diffraction envelope, opening up the possibility of a lateral force on the electron [7]. Reversing the magnetic-field direction in the solenoid reverses the direction of the asymmetry, as has also been measured by Becker et al. These are significant results, experimentally demonstrating that the dominant assumption of zero forces in the AB effect [16] may be incorrect.

Here, we see that Eq. (17) has a $\sin^2\theta$ and $\cos^2\theta$ term which angle-average over the top or bottom of the solenoid to a non-zero value of 0.5, and a directionality that is proportional to the sign of the **B** · d**S** dot product determining $\Phi_B$. The $\sin^2\theta$ dependence cancels the $\cos^2\theta$ term only at specific radii and angles, but not over the entire field (see Appendix B); our results are thus consistent with the experiments of Becker et al. The circulation of the vector potential (i.e., the magnetic flux $\Phi_B$) therefore creates a lateral force $F_y$. This is somewhat analogous to the Magnus effect in hydrodynamics, where the velocity circulation around a spinning cylinder which is also moving with an axial velocity results in a lateral force on the cylinder.

Pozzi et al. have also published detailed results of such an experiment [16]. While they use the results to conclude that there are no lateral forces, their supplementary material indicates that there is in fact some deflection of the electron envelope with respect to the fringe pattern. Specifically, they report in Fig. S7 of their Supplementary Material a fringe shift of 10.5 pixels for an envelope shift of 2.8 pixels. Our results are generally consistent with a deflection of the envelope.

Finally, the third class of experiments against which our theory can be compared is that of the phase shift $\Delta\phi_{AB} = e\Phi_B/\hbar$. This shift has been derived in many AB papers using a quantum-mechanical approach [1]; in Appendix C, we obtain the same result using a semi-classical derivation that confirms our use of Laplace's equation for the electron velocity field and the conservation of canonical momentum $m\Delta\mathbf{v} = -e\Delta\mathbf{A}$ in the azimuthal direction where $\partial A_\theta(r)/\partial\theta = 0$, giving $|\Delta\mathbf{A}| = 2A_\theta = \Phi_B/\pi R$. It is also consistent with the quantum mechanical approach of Ref. [10] showing that the AB phase accumulates locally.

## IV. Summary and Conclusions

We have shown that the shear of the vector potential in the AB effect leads to the possibility of a classical force on an electron as it propagates past the solenoid. This force is a result of the electron traveling through the gradient $\nabla \mathbf{A}$ of the vector potential; while it is similar in *form* to the velocity-dependent "*evB*" Lorentz force, it is not a Lorentz force, as there is no curl of the vector potential in the region outside the solenoid. Instead, we have identified this force to be a result of the shear of $\mathbf{A}$, rather than its curl or divergence.

Our results are consistent with all known classes of AB experiments, including a phase shift which produces a measureable interference fringe shift, the absence of time delays along the direction of propagation, and the possibility of lateral forces deflecting the electron.

Note that the derivation reported here is not a synthetic field, and is distinct from the quantum approach recently taken by Aharonov's collaborators on the possibility of magnetic forces in the absence of a conventional ($\nabla \times \mathbf{A}$) magnetic field [17]. The introduction of vector-potential shear to the AB problem thus opens up the possibility of a classical (local) force in regions of space where there are neither electric nor magnetic fields.

## Appendix A – Convective Momentum Transport

In this Appendix, we derive the expression for the convective derivative term $(\mathbf{v} \cdot \nabla)\mathbf{A}$ in the cylindrical-coordinate geometry used for the AB effect. With $\mathbf{A} = A_\theta(r)\hat{\boldsymbol{\theta}}$, we have

$$(\mathbf{v}\cdot\nabla)\mathbf{A} = \left[(v_r\hat{\mathbf{r}} + v_\theta\hat{\boldsymbol{\theta}})\cdot\left(\frac{\partial}{\partial r}\hat{\mathbf{r}} + \frac{1}{r}\frac{\partial}{\partial\theta}\hat{\boldsymbol{\theta}}\right)\right]A_\theta(r)\hat{\boldsymbol{\theta}}$$
$$= \left[v_r\frac{\partial(A_\theta\hat{\boldsymbol{\theta}})}{\partial r} + \frac{v_\theta}{r}\frac{\partial(A_\theta\hat{\boldsymbol{\theta}})}{\partial\theta}\right]$$
(A1)

which, by expanding the partial derivatives and using $\partial\hat{\boldsymbol{\theta}}/\partial r = 0$ and $\partial\hat{\boldsymbol{\theta}}/\partial\theta = -\hat{\mathbf{r}}$, we obtain

$$\frac{d\mathbf{A}}{dt} = (\mathbf{v}\cdot\nabla)\mathbf{A} = v_r\frac{\partial A_\theta}{\partial r}\hat{\boldsymbol{\theta}} - \frac{v_\theta A_\theta}{r}\hat{\mathbf{r}} \quad (A2)$$

for the total derivative $d\mathbf{A}/dt$. Note that it is the radial velocity which determines the tangential component of the force. We also see a radial component of the force, analogous to the $v_\theta^2/r$ term in the hydrodynamic derivation of $(\mathbf{v}\cdot\nabla)\mathbf{v}$. This radial term is not a shear force; it is a centripetal force that drives the change in direction of the electron velocity vector.

## Appendix B – Lateral Forces

In this Appendix, we show that the net lateral force over the entire field in Eq. (17) must be greater than zero. We start with the lateral force distribution of Eq. (17), which is zero only for an angle $\theta_o$ where

$$\tan^2\theta_o = \frac{1 - R^2/r^2}{1 + R^2/r^2} \qquad (B1)$$

for $r \geq R$. For a given $R/r \leq 1$, there are only two angles $\pm\theta_o$ (and symmetric negative angles, as defined in Fig. 2) at which the lateral force is zero (Fig. B1). In addition, the $1+R^2/r^2$ factor for the $\sin^2\theta$ term guarantees that its peak magnitude is always larger than that of the $\cos^2\theta$ term, except at $r$

= ∞ where they are equal. Remembering that $F_y$ is a force *distribution*, we see that the net force obtained by integrating over the entire field consisting of the $\sin^2\theta$ and $\cos^2\theta$ terms must be greater than zero.

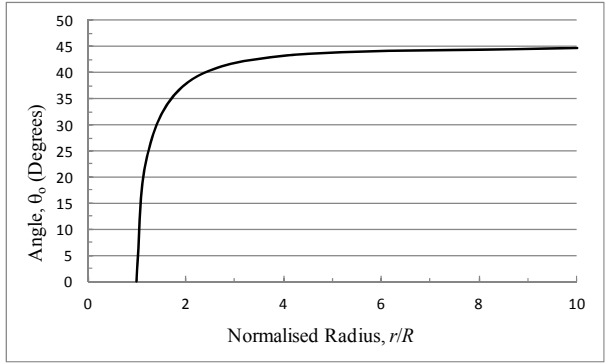

Figure B1 – Plot of the angle $\theta_o$ at which the lateral force in Eq. (17) is zero for a given $r/R$ outside the solenoid.

## Appendix C – Semi-Classical Derivation of the AB Phase Shift

In this Appendix, we derive the expression for the AB phase shift $\Delta\phi_{AB} = e\Phi_B/\hbar$ using Laplace's equation for the velocity field and the conservation of canonical momentum $m\Delta\mathbf{v} = -e\Delta\mathbf{A}$ in the azimuthal direction where $\partial A_\theta(r)/\partial\theta = 0$. These illustrate the difference in tangential velocity across the upper and lower halves of the solenoid required for the fringe shift. This is shown in Fig. C1, where the circulation term in Eq. (10) results in an asymmetry in the tangential velocity distribution, with a higher velocity on the top of the solenoid than the bottom for a clockwise velocity circulation (or counter-clockwise vector-potential circulation, as shown in Fig. 1). This determines the difference in de Broglie wavelengths, and thus the phase difference between the upper and lower halves, as shown heuristically in Ref. [9].

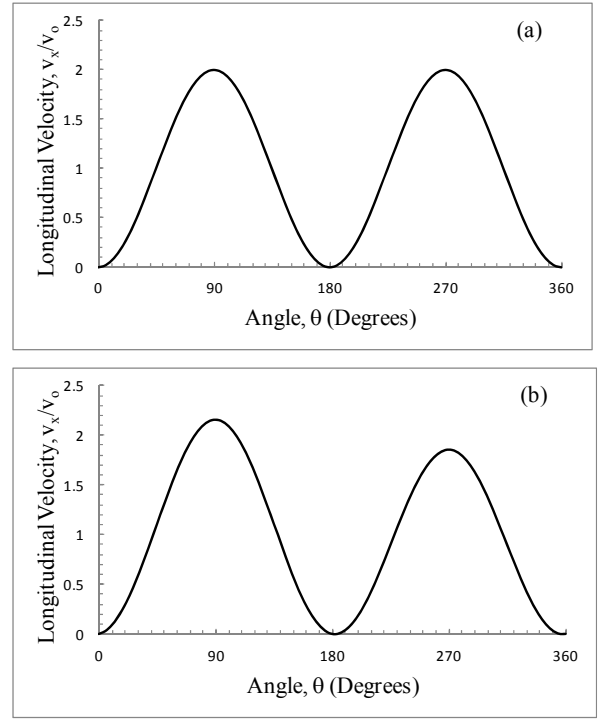

Fig. C1 – Angular dependence of the normalized *x*-component of the electron velocity at $r = R$ where $v_r = 0$ and $v_x = -v_\theta\sin\theta$. Shown is a comparison of (a) the velocity symmetry across the top half ($\theta = 180 \rightarrow 0$ degrees) and bottom half ($\theta = 180 \rightarrow 360$ degrees) of the solenoid when $\Phi_B = 0$; and (b) the asymmetry with $\Phi_B > 0$. In practice, the difference in normalized peak velocities will be on the order of $2eA_\theta/mv_o \approx$ 1 part in $10^6$ for $v_o = 0.6 \times 10^8$ m/s ($E_o = 10$ keV).

More specifically, the phase of the electron de Broglie wave is given by $\phi = \mathbf{k}\cdot\mathbf{s}$. By choosing a trajectory in the azimuthal direction around the solenoid at a fixed radius ($r = R$) where $A_\theta$ is constant, we find the phase difference $\Delta\phi$ between the top and bottom halves of the solenoid

$$\Delta\phi = \mathbf{s}\cdot\Delta\mathbf{k} = r\cdot\Delta k_r + r\theta_t\cdot\Delta k_\theta \quad (C1)$$
$$= R\theta_t\cdot\Delta k_\theta$$

for a path difference $\Delta\mathbf{s} = 0$, a traversed angle $\theta_t$ around the solenoid, and a difference in wavenumber $\Delta k_\theta$ based on the difference in tangential velocity $v_\theta$ at $r = R$ where $v_r = 0$. Writing out the wavenumbers for the electron for $\psi_1$ and $\psi_2$ at the top and bottom of the solenoid, we have

$$k_{\theta 1} = \frac{2\pi}{\lambda_1} = \frac{m|v_{\theta 1}|}{\hbar} \text{ and } k_{\theta 2} = \frac{2\pi}{\lambda_2} = \frac{m|v_{\theta 2}|}{\hbar} \quad (C2)$$

where the tangential velocity [Eq. (10)] and the velocity circulation term $\Gamma_v$ [Eq. (11)] combine to give

$$|v_{\theta 1}| = \left| 2v_o \sin\theta_1 + \frac{e}{m}\frac{\Phi_B}{2\pi R} \right| \quad (C3)$$

and

$$|v_{\theta 2}| = \left| 2v_o \sin\theta_2 + \frac{e}{m}\frac{\Phi_B}{2\pi R} \right| \quad (C4)$$

Substituting these results in Eq. (C2), we obtain from Eq. (C1)

$$\Delta\phi = \pi R \cdot \Delta k_\theta = R\theta_t \frac{m}{\hbar}\left[|v_{\theta 1}| - |v_{\theta 2}|\right] \quad (C5)$$

Starting at the leading edge of the solenoid where $\theta_1 = \theta_2 = \pi$ rads, and propagating to the trailing edge where $\theta_1 = 0$ and $\theta_2 = 2\pi$ radians, we evaluate Eq. (C5) numerically, obtaining a constant value for the difference in azimuthal velocities everywhere except at the points for the leading and trailing edges (where a difference cannot be defined). That constant depends on the velocity-circulation term in Eq. (11)

$$|v_{\theta 1}| - |v_{\theta 2}| = \frac{\Gamma_v}{\pi R} = -\frac{e}{m}\frac{\Phi_B}{\pi R} = -\frac{e}{m}\Delta\mathbf{A} \quad (C6)$$

which, when combined with Eq. (C5) and $\theta_t = \pi$ rads for $\psi_1$ and $\psi_2$, gives us the well-known non-dispersive expression $\Delta\phi_{AB} = e\Phi_B/\hbar$ for the AB phase shift. This semi-classical derivation confirms our use of Laplace's equation and conservation of canonical momentum $m\Delta\mathbf{v} = -e\Delta\mathbf{A}$ in the azimuthal direction, where $|\Delta\mathbf{A}| = 2A_\theta = \Phi_B/\pi R$; it is also consistent with the quantum mechanical results of Ref. [10] showing that the AB phase accumulates locally along the path.

**Notes and References**